\documentclass[superscriptaddress,twocolumn,floatfix,aps]{revtex4-2}
\usepackage[utf8]{inputenc}
\usepackage{xcolor}
\usepackage{bbm}
\usepackage{bm}
\usepackage{physics}
\usepackage{graphicx}
\usepackage{placeins}
\usepackage{float}
\usepackage{chronology}
\usepackage{amsmath,amsfonts,amssymb,amsthm}
\usepackage{qcircuit}
\usepackage{indentfirst}
\usepackage{dsfont}
\usepackage[normalem]{ulem}
\usepackage{graphicx,wasysym}
\usepackage{mathtools}
\usepackage{hyperref}
\usepackage{tcolorbox}
\usepackage{booktabs}

\newcommand{\sgn}[1]{\text{sgn}\left\lbrack#1\right\rbrack}

\newcommand{\xtilde}{\tilde{\mathbf{x}}}
\newcommand{\xj}{\mathbf{x}_j}

\begin{document}
\title{Compact quantum kernel-based binary classifier}

\author{Carsten Blank}
\affiliation{data cybernetics ssc GmbH, Landbergs am Lech, Germany}
\author{Adenilton J. da Silva}
\affiliation{Centro de Inform\'{a}tica, Universidade Federal de Pernambuco, Recife, Pernambuco, Brazil}
\author{Lucas P. de Albuquerque}
\affiliation{Centro de Inform\'{a}tica, Universidade Federal de Pernambuco, Recife, Pernambuco, Brazil}
\author{Francesco Petruccione}
\affiliation{School of Chemistry and Physics, University of KwaZulu-Natal, Durban, South Africa}
\affiliation{National Institute for Theoretical and Computational Sciences (NITheCS), South Africa}
\affiliation{School of Data Science and Computational Thinking and Department of Physics, Stellenbosch University, Stellenbosch 7602, South Africa}
\author{Daniel K. Park}
\email{dkd.park@yonsei.ac.kr}
\affiliation{Department of Applied Statistics, Yonsei University, Seoul 03722, Korea}
\affiliation{Department of Statistics and Data Science, Yonsei University, Seoul 03722, Korea}

\begin{abstract}
Quantum computing opens exciting opportunities for kernel-based machine learning methods, which have broad applications in data analysis. Recent works show that quantum computers can efficiently construct a model of a classifier by engineering the quantum interference effect to carry out the kernel evaluation in parallel. For practical applications of these quantum machine learning methods, an important issue is to minimize the size of quantum circuits. We present the simplest quantum circuit for constructing a kernel-based binary classifier. This is achieved by generalizing the interference circuit to encode data labels in the relative phases of the quantum state and by introducing compact amplitude encoding, which encodes two training data vectors into one quantum register. When compared to the simplest known quantum binary classifier, the number of qubits is reduced by two and the number of steps is reduced linearly with respect to the number of training data. The two-qubit measurement with post-selection required in the previous method is simplified to single-qubit measurement. Furthermore, the final quantum state has a smaller amount of entanglement than that of the previous method, which advocates the cost-effectiveness of our method. Our design also provides a straightforward way to handle an imbalanced data set, which is often encountered in many machine learning problems.
\end{abstract}
\maketitle

\section{Introduction}

As the quest for fault-tolerant quantum computers continues, noisy intermediate-scale quantum (NISQ) computers are expected to be available in the near future~\cite{Preskill2018quantumcomputing}, supported by the recent technological advances in quantum computing~\cite{doi:10.1098/rsta.2011.0352,doi:https://doi.org/10.1002/9781118742631.ch08,Park_Kyungdeock2015,Google_QS,ion_trap_BM,HF_google,Zhong1460,harrigan_quantum_2021}. Although the size of quantum circuits that NISQ devices can execute reliably is limited, the size of the quantum state space efficiently manipulated by them is much beyond what classical computers can handle. As such an interesting era is within reach, an important task in the quantum computing community is finding commercially relevant applications of the NISQ technology for which quantum advantage can be demonstrated in the near future.

Machine learning has been considered as a promising domain for which quantum computing can shine~\cite{wittek,QML-Biamonte,SupervisedQML,Dunjko_2018,QML_PRSA,schuld2021quantum}. Quantum advantages in machine learning are expected, since quantum computers can in principle store and manipulate the amount of classical information that scales exponentially with the number of qubits~\cite{PhysRevLett.100.160501,ffqram,9259210}. Moreover, quantum computers can reduce the computational cost exponentially for solving certain basic linear algebra problems~\cite{Nielsen:2011:QCQ:1972505,PhysRevLett.103.150502_HHL_qBLAS} that often appear as basic subroutines in machine learning tasks, such as in support vector machine~\cite{PhysRevLett.113.130503_qSQVM} and principal component analysis~\cite{qPCA}. However, the size of quantum circuits required for implementing basic linear algebra subroutines on a quantum computer is too large for near-term quantum devices. 

Several quantum machine learning algorithms have been proposed to perform the kernel-based classification by exploiting the ability of quantum computers to efficiently evaluate inner products in an exponentially-large Hilbert space~\cite{schuld2021quantum}, and without relying on expensive subroutines ~\cite{Havlicek2019,PhysRevLett.113.130503_qSQVM,PhysRevLett.122.040504,htc,blank_quantum_2020,PARK2020126422}. In particular, Refs.~\cite{htc,blank_quantum_2020} proposed the most simple quantum circuit for utilizing the quantum interference effect now known as the Hadamard-test classifier (HTC) and show that it can be used as a simple model of a kernel-based classifier for real-valued data. If an efficient state preparation routine is known, the algorithm achieves the logarithmic scaling in the dimension and number of the input data with a simple setup. Namely, given a quantum state that encodes the classical data in a specific form, the algorithm only uses a Hadamard gate and the expectation measurement of a two-qubit observable to complete the labeling task. Furthermore, the algorithm is agnostic to the quantum data encoding method, such as amplitude encoding~\cite{SupervisedQML} and quantum feature mapping~\cite{PhysRevLett.122.040504}.

In this work, we present a kernel-based quantum binary classifier that is even simpler than HTC by introducing compact amplitude encoding (CAE) of real-valued data, which reduces the number of training steps linearly and the number of qubits by two. In CAE, training data belonging to one class is encoded as the real part of the probability amplitudes, while the remaining training data is encoded as the imaginary part. In order to utilize CAE, we show that the single-qubit interfering circuit of the HTC can be generalized to take the imaginary part of the quantum state into account. In this way, the label information encoding is not explicitly executed on a quantum circuit and two sets of data are encoded in a single quantum register, thereby eliminating the state preparation subroutine for preparing the label registers and reducing the number of index qubits. Furthermore, our classifier provides a simple method for assigning arbitrary weights to two training data sets with different labels, which broadens the application of our method to imbalanced data sets in which the number of training data points in two classes are unequal. The CAE also localizes data so that the entanglement is reduced compared to the HTC. Although this finding is only on a numerical basis, we suspect that this is a general feature. This could lead to better performance of the classifier in the NISQ era since less entanglement implies reduced circuit complexity~\cite{MORA2006,PhysRevLett.95.200503,2110.13454,2111.03132}. 

The remainder of the paper is organized as follows. Section~\ref{sec:preliminaries} provides theoretical backgrounds and reviews for this study, such as the description of the classification problem and the existing kernel-based quantum classifier that has been known to be simplest. Section~\ref{sec:algorithm} explains the main classification algorithm proposed in this work. In Sec.~\ref{sec:resource}, we carry out the entanglement analysis with numerical simulations of classification on Iris and Wine data sets as examples. The simulation results show that the classifier proposed in this work is more compact than the HTC with respect to the amount of entanglement the circuit produces. Section~\ref{sec:conclusion} concludes and discusses future research directions.

\section{Quantum kernel-based classifier}
\label{sec:preliminaries}
\subsection{Binary classification}

Classification is a canonical pattern recognition problem that aims to label a data point $\tilde{x} \in \mathbbm{R}^N$ as accurately as possible given a labelled (or training) data set
$$\mathcal{D} = \left\{ (x_0, y_0), \ldots, (x_{M-1}, y_{M-1}) \right\} \subset \mathbbm{R}^N\times\{0,1,\ldots,L-1\}.$$ 
Since the training data set includes labels, this task can be addressed via a supervised machine learning technique. Among many machine learning approaches, a kernel method provides a straightforward interpretation of the classification process. It is based on choosing a feature space for the data set such that the classification score is defined as a linear function of the similarity (i.e. kernel) between each training data  and the test data. This principle naturally connects quantum computing to kernel-based classification since the quantum Hilbert space can be used as the data feature space with a proper definition of the similarity~\cite{PhysRevLett.122.040504,schuld2021quantum}.

In this work, we focus on real-valued data as is common in practical machine learning tasks. Moreover, we focus on binary classification (i.e. $L=2$) since a multi-class classification can be constructed with binary classifiers by one versus all or one versus one scheme~\cite{giuntini2021quantum}. Note that in binary classification,  the two class labels are often denoted as $\pm 1$.

\subsection{Amplitude encoding}
The first step towards utilizing the quantum Hilbert space as the feature space is encoding classical data as a quantum state. Although the best encoding strategy remains an open problem, many previous quantum machine learning algorithms chose to represent a classical vector $\mathbf{x}_j = (x_{0j}, \ldots, x_{(N-1)j})^T \in \mathbbm{R}^N$ as probability amplitudes of a quantum state in the following form~\cite{PhysRevLett.113.130503_qSQVM,qPCA,htc,SupervisedQML,qcnn},
\begin{equation}
\label{eq:amplitude_encoding_1}
    \ket{\mathbf{x}_j} := \sum_{i=0}^{N-1} x_{ij} \ket{i},
\end{equation}
using $\lceil \log_2(N)\rceil$ qubits, where the input vector is normalized and have unit length, i.e. $\|\mathbf{x}_j\|=1$. The above form of data encoding is often called amplitude encoding, and it can be generalized for a set of $M$ data points $\mathbf{x}_1, \ldots, \mathbf{x}_M$ as
\begin{equation}
\label{eq:amplitude_encoding_M}
    \frac{1}{\sqrt{M}} \sum_{j=0}^{M-1}\sum_{i=0}^{N-1} x_{ij} \ket{i}\otimes\ket{j},
\end{equation}
which uses at least $\lceil \log_2(NM)\rceil$ qubits.

Hereinafter, we will omit the Kronecker product symbol ($\otimes$) and write $\ket{i}\otimes\ket{j}= \ket{ij}$ whenever the meaning is clear.

\subsection{Hadamard-test classifier}

A simple model of quantum classifier for real-valued data was introduced in Ref.~\cite{htc}, and is nowadays referred to as the Hadamard-test classifier (HTC). This work presents improvements of the HTC in several aspects, and hence this section briefly reviews the algorithm's structure.

For the Hadamard-test classifier, the data set $\mathcal{D}$ is encoded in a quantum state as
\begin{equation}
\label{eq:htc_initial}
	\ket{\psi} = \frac{1}{\sqrt{2}} \sum_{j=0}^{M-1} \sqrt{a_j}\left(\ket{0}\ket{\mathbf{x}_j} + \ket{1}  \ket{\tilde{\mathbf{x}}} \right)\ket{y_j}\ket{j},
\end{equation}
where $\ket{\mathbf{x}_j}$ and $\ket{\tilde{\mathbf{x}}}$ are quantum representatives of the  classical training and test data vectors by utilizing an encoding of choice (a feature map). The label $y_j\in\lbrace -1,+1 \rbrace$ is transformed to the computational basis of the label qubit with the rule $y_j\rightarrow |(1-y_j)/2\rangle\in\lbrace |0\rangle,|1\rangle \rbrace$. In the original paper~\cite{htc}, the weights were chosen uniform, i.e. $a_j = 1/M\;\forall j$, while it has been shown~\cite{PARK2020126422} that it can be used as a variable to be optimized, similar to the treatment in support vector machines. Principally, the HTC applies a Hadamard-test as  measurement scheme: this applies a Hadamard gate to the ancillary qubit in order to interfere training and test data states. This is finalized by measuring an expectation value of a two-qubit observable $\sigma_z^{(a)}\sigma_z^{(l)}$ on the ancillary and label qubit, leading up to
\begin{equation}
	\label{eq:htc_score}
    \langle\psi|H^{(a)} \sigma_z^{(a)}\sigma_z^{(l)}H^{(a)}|\psi\rangle=\sum_{j=0}^{M-1}a_jy_j\braket{\tilde{\mathbf{x}}}{\mathbf{x}_j}.
\end{equation}
The superscripts $a$ and $l$ are the qubits on which the corresponding operator is applied, namely the ancillary and the label qubit, respectively. The connection between the HTC and the field of kernel methods is based on this equation. One can see that the kernel function is $k(\mathbf{x}_j,\tilde{\mathbf{x}})=\braket{\mathbf{x}_j}{\tilde{\mathbf{x}}}$. Consequently, the right-hand side of Eq.~(\ref{eq:htc_score}) defines the classification score, which is denoted by $f$. Therefore, HTC assigns a new label to the test data by the rule 
\begin{equation}
\label{eq:classifier}
\tilde{y} = \sgn{\sum_{j=0}^{M-1}a_jy_j\braket{\mathbf{x}_j}{\tilde{\mathbf{x}}}}.
\end{equation}

\section{Compact classifier}
\label{sec:algorithm}

\subsection{Generalization}
The generalization of the interference circuit of the HTC was discussed in Ref.~\cite{park2021robust} as a means to show that the Hadamard gate is the optimal choice for minimizing the number of sampling. In this work, we take a step further and utilize the generalized interference circuit to reduce the quantum circuit cost of the HTC.

The generalized interference circuit for the HTC uses $R_z(\phi)R_y(\theta_0)$ in place of the first Hadamard gate, and uses $R_y(\theta_1)$ in place of the last Hadamard gate, where  $R_z(\phi)=\cos(\phi/2)I-i\sin(\phi/2)\sigma_z$ and  $R_y(\theta)=\cos(\theta/2)I-i\sin(\theta/2)\sigma_y$ are the single-qubit rotation gates. As shown in Ref.~\cite{park2021robust}, the two-qubit expectation value measured in the generalized interference circuit is
\begin{align}
\label{eq:expval_general1}
    \langle\sigma_{z}^{(a)}\sigma_z^{(l)}\rangle = & \sum_{j=0}^{M-1}a_jy_j\big{(}\cos(\theta_0)\cos(\theta_1) -\sin(\theta_0)\sin(\theta_1)\nonumber\\
    & \times \left(\cos(\phi)\text{Re}\braket{\tilde{\mathbf{x}}}{\mathbf{x}_j}-\sin(\phi)\text{Im}\braket{\tilde{\mathbf{x}}}{\mathbf{x}_j}\right)\big{)}.
\end{align}
While the original HTC only uses the real part of the inner product (i.e., $\ip{\xtilde}{\xj}$) in the classification algorithm, the generalized interference circuit opens up the possibility of harnessing the imaginary part as well. Since the goal of this paper is to utilize the imaginary part, we set $\theta_0=\pi/2$ and $\theta_1=-\pi/2$ for simplicity. In this case, Eq.~(\ref{eq:expval_general1}) becomes
\begin{align}
\label{eq:expval_general2}
    \langle\sigma_{z}^{(a)}\sigma_z^{(l)}\rangle = \sum_{j=0}^{M-1}a_jy_j \left(\cos(\phi)\text{Re}\braket{\tilde{\mathbf{x}}}{\mathbf{x}_j}-\sin(\phi)\text{Im}\braket{\tilde{\mathbf{x}}}{\mathbf{x}_j}\right)\big{)}.
\end{align}
The above result can also be obtained by applying a single-qubit rotation gate $R_z(\phi)$ to the ancilla qubit of the HTC in Eq.~(\ref{eq:htc_initial}). 

\subsection{Compact amplitude encoding}

The main idea in this work is based on the observation that if $\cos(\phi)$ and $\sin(\phi)$ in Eq.~(\ref{eq:expval_general2}) have the same sign, then the real and imaginary parts of the state overlap contributes with opposite sign. Thus by encoding the training data with label $+1$ ($-1$) to real (imaginary) part of the probability amplitudes of the quantum state, the binary classification can be done without explicitly preparing the label register. This leads to reducing the number of qubits by one. Note that when the number of training data vectors in two classes is the same, $\cos(\phi)=\sin(\phi)=1/\sqrt{2}$. In Sec.~\ref{sec:III.C}, we discuss how to control $\phi$ for imbalanced data set.

With this background, the compact amplitude encoding (CAE) is introduced to utilize the imaginary part as follows. In CAE, two $N$-dimensional real vectors $\mathbf{x}_k^{+} = (x_{0k}^{+}, \ldots, x_{(N-1)k}^{+})^T $ and $\mathbf{x}_k^{-} = (x_{0k}^{-}, \ldots, x_{(N-1)k}^{-})^T $ with the corresponding labels indicated by the superscript ($\pm$) are loaded in a quantum state in the following form,
\begin{equation}
\label{eq:CAE}
    \ket{\mathbf{x}_k}_c := \sum_{j=0}^{N-1} (x_{jk}^{+}+ix^{-}_{jk}) \ket{j},
\end{equation}
where $\|\mathbf{x}_{k}^{+}\|^2+\|\mathbf{x}^{-}_{k}\|^2 = 1$ to satisfy the normalization condition. Note that various scaling methods can be employed to satisfy this condition, and one of the natural ways is to scale each vector so that $\|\mathbf{x}_k^{\pm}\|=1/\sqrt{2}$. We implicitly assume this way of normalizing vectors unless stated otherwise. The above equation shows that two $N$-dimensional vectors are encoded in $\lceil \log_2(N)\rceil$ qubits. The subscript $c$ in Eq.~(\ref{eq:CAE}) distinguishes the quantum state from amplitude encoding, and indicates that the classical vector is encoded via CAE. It is important to note that $|\cdot\rangle_c$ means two data points are encoded in the state vector, whereas the ordinary ket vector without the subscript $c$ means one data point is encoded. In addition, we define two more states as
\begin{equation}
    |\mathbf{x}_k^{\pm}\rangle := \frac{1}{\|\mathbf{x}_k^{\pm}\|}\sum_{j=0}^{N-1} x_{jk}^{\pm} \ket{j}.
\end{equation}
Then the state overlap between two quantum registers, one encodes an $N$-dimensional real vector $\tilde{\mathbf{x}}$ via amplitude encoding (Eq.~(\ref{eq:amplitude_encoding_1})) and the other encodes two $N$-dimensinal real vectors $\mathbf{x}_k^+$ and $\mathbf{x}_k^-$ via CAE (Eq.~(\ref{eq:CAE})) is
\begin{equation}
\label{eq:overlap_CAE}
    \langle \tilde{\mathbf{x}} | \mathbf{x}_k\rangle_c = \frac{1}{\sqrt{2}}\left(\braket{\tilde{\mathbf{x}}}{\mathbf{x}_k^{+}}+i\braket{\tilde{\mathbf{x}}}{\mathbf{x}_k^{-}} \right).
\end{equation}

\subsection{Compact quantum binary classifier}
\label{sec:III.C}

\begin{figure*}[t]
    \centering
    \includegraphics[width=0.98\textwidth]{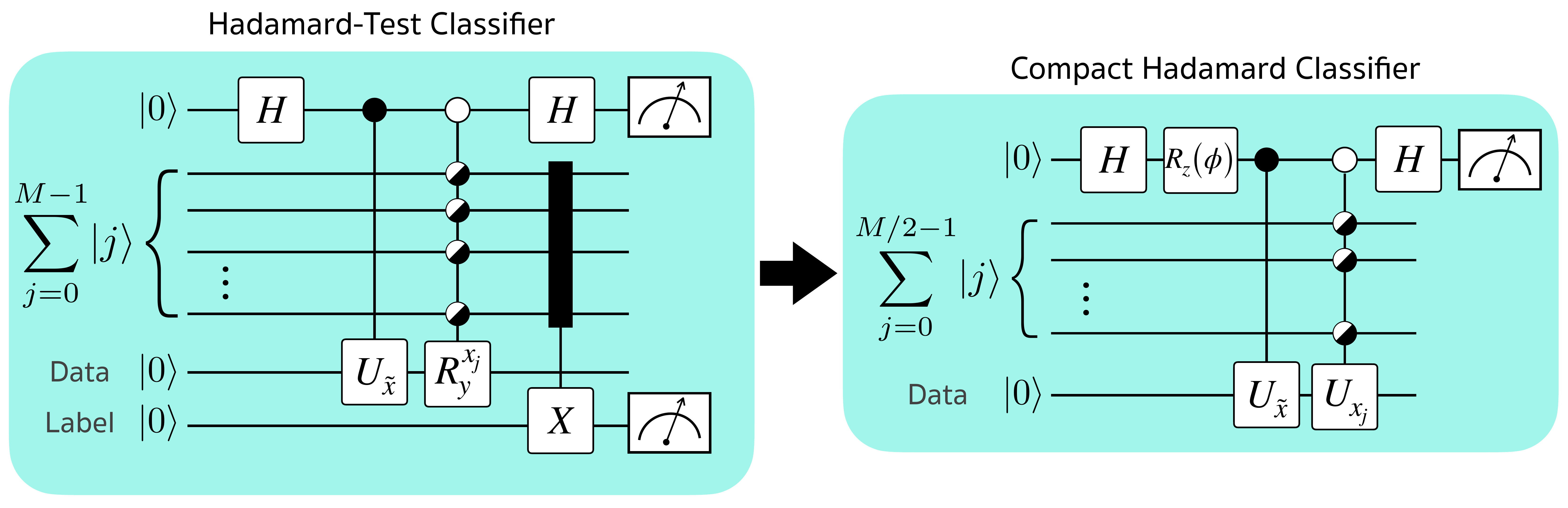}
    \caption{The comparison of quantum circuits for the Hadamard-test classifier (left) and compact Hadamard classifier (right). The half-filled circles indicate that the unitary operation is the uniformly controlled gate~\cite{Mottonen:2005:TQS:2011670.2011675,bergholm2005quantum}. For the Hadamard-test classifier, the filled rectangle connected with a vertical line to X applied to the label qubit indicates that the number of control qubits depends on the imbalance in the label proportion.}
    \label{fig:circuits}
\end{figure*}

The compact quantum machine learning algorithm that implements the binary classifier from Eq.~(\ref{eq:classifier}) is constructed as follows. We first prepare an initial state that encodes the data set $\mathcal{D}$ as
\begin{equation}
\label{eq:initial_compact}
    	\ket{\psi_i} = \frac{1}{\sqrt{2}} \sum_{j=0}^{\frac{M}{2}-1} \sqrt{b_j}\left(\ket{0}\ket{\mathbf{x}_j}_c + e^{-i\phi}\ket{1}  \ket{\tilde{\mathbf{x}}} \right)\ket{j},
\end{equation}
where the subscript $c$ indicates that the state is prepared via CAE. It is important to note that $\sum_{j=1}^{M/2} b_j = 1$, and hence the set of weights is different from that of the Hadamard-test classifier which satisfies $\sum_{j=0}^{M-1}a_j = 1$. For example, for a set of uniform weights, $b_j = 2/M$ and $a_j = 1/M$. For simplicity, we assume that the number of data with label $+1$ denoted by $M_{+}$ is equal to the number of data with label $-1$ denoted by $M_{-}$. The state above is easier to prepare than the state required in HTC shown in Eq.~(\ref{eq:htc_initial}) since the label information is not explicitly encoded and the relative phase $e^{-i\phi}$ can be added by applying a single-qubit rotation gate $R_z(\phi)$ on the ancilla qubit. Moreover, since two training data are encoded in one quantum register, the number of terms in the summation is decreased by a factor of 2, meaning that the dimension of the index register is also decreased by a factor of 2. After the state preparation, the rest of the algorithm only requires a Hadamard gate and the measurement of the ancilla qubit in the $\sigma_z$ basis. The Hadamard gate interferes the copies of the new input and the training inputs to produce a state
\begin{align}
\label{eq:final_compact}
    	\ket{\psi_f} = \frac{1}{2} \sum_{j=0}^{\frac{M}{2}-1} \sqrt{b_j}\big{[}&\ket{0}(\ket{\mathbf{x}_j}_c +e^{-i\phi}\ket{\tilde{\mathbf{x}}})\nonumber \\
    	+&\ket{1} (\ket{\mathbf{x}_j}_c-e^{-i\phi}\ket{\tilde{\mathbf{x}}}) \big{]}\ket{j}.
\end{align}
The probability to measure the ancilla qubit in $|0\rangle$ is given as
$$
    \Pr(0) = \frac{1}{4}\sum_{j=0}^{\frac{M}{2}-1} b_j\left(2+e^{-i\phi}\textsubscript{c}\langle \mathbf{x}_j|\tilde{\mathbf{x}}\rangle+e^{i\phi}\langle\tilde{\mathbf{x}}|\mathbf{x}_j\rangle_c\right).
$$
By denoting $\kappa_j=\langle\tilde{\mathbf{x}}|\mathbf{x}_j\rangle_c$, the above equation can be written as
\begin{equation}
    \Pr(0) = \frac{1}{2}\sum_{j=0}^{\frac{M}{2}-1} b_j\left(1+\cos(\phi)\text{Re}(\kappa_j)-\sin(\phi)\text{Im}(\kappa_j)\right).
\end{equation}
Similarly, the probability to measure the ancilla qubit in $|1\rangle$ can be calculated as
\begin{equation}
    \Pr(1) = \frac{1}{2}\sum_{j=0}^{\frac{M}{2}-1} b_j\left(1-\cos(\phi)\text{Re}(\kappa_j)+\sin(\phi)\text{Im}(\kappa_j)\right).
\end{equation}
Therefore, the expectation value of the $\sigma_z$ operator measured on the ancilla qubit is
\begin{equation}
\label{eq:expval_chtc}
    \langle \sigma_z^{(a)}\rangle = \sum_{j=0}^{\frac{M}{2}-1}b_j\left(\cos(\phi)\text{Re}(\kappa_j)-\sin(\phi)\text{Im}(\kappa_j)\right).
\end{equation}
By setting $\cos(\phi)=\sin(\phi)=1/\sqrt{2}$ and using Eq.~(\ref{eq:overlap_CAE}), we arrive at
\begin{equation}
    \langle \sigma_z^{(a)}\rangle = \frac{1}{2}\sum_{j=0}^{\frac{M}{2}-1}b_j\left(\langle\tilde{\mathbf{x}}|\mathbf{x}_j^{+}\rangle - \langle\tilde{\mathbf{x}}|\mathbf{x}_j^{-}\rangle\right).
\end{equation}
Since the state overlap for the training data in class $+1$ ($-1$) always have the positive (negative) sign, the above equation can be written in its final form as
\begin{align}
\label{eq:chtc_final}
    \langle \sigma_z^{(a)}\rangle = \frac{1}{2}\sum_{j=0}^{M-1}b'_jy_j\langle \tilde{\mathbf{x}}|\mathbf{x}_j\rangle,
\end{align}
where $b'_{m} = b'_{m+M/2} = b_m$ for $m=0,\ldots,M/2-1$ and $\sum_{j=0}^{M-1}b'_j = 2$. This outcome is very similar to the classification score in the Hadamard classifier obtained by the two-qubit measurement. The constant factor of $1/2$ is attributed to the different normalization conditions between the set of weights $b_j$ and $a_j$ as described below Eq.~(\ref{eq:initial_compact}). We refer to this classifier as compact Hadamard classifier (CHC). The CHC is obtained with a single-qubit measurement implying that one can bypass the standardization of the training data set required in HTC for increasing the post-selection probability. The comparison of quantum circuits for implementing HTC and CHC is depicted in Fig.~\ref{fig:circuits}.

The reduction of quantum circuit sizes, which is of critical importance in practice, achieved with CHC is as follows. By introducing a controllable relative phase between two computational basis states of the ancilla qubit and using compact amplitude encoding, the number of qubits is reduced by two, one for the label register and another in the index register. Then the quantum operations for encoding the training data set is reduced from $M$ gates controlled by $\log_2(M)$ index qubits to $M/2$ gates controlled by $\log_2(M)-1$ index qubits. Having one fewer controlled qubit can further reduce the quantum circuit depth by a factor of two~\cite{PhysRevA.52.3457}. Therefore, the number of operations for encoding the training data set $\mathbf{x}_j$ is reduced by a factor of four. The number of operations in CHC is also reduced due to the removal of operations for explicitly encoding the label information in a separate register. The number of gates needed for preparing the label register in HTC is 1 when the number of data belonging to each class is equal since one controlled-NOT (CNOT) gate controlled by one of the index qubits applied to the label qubit can split the Hilbert space into two subspaces with an equal number of 0's and 1's in the label qubit. But the number of operation increases linearly with the difference in the number of data with different labels. For example, if $\alpha M$ and $\beta M$ data belong to class $+1$ and $-1$, respectively, where $\alpha+\beta =1$, then the number of operations necessary to encode label information grows as $|\alpha-\beta|M$. Therefore, in total, our compact classifier can reduce the number of operations at least by a factor of four, while the reduction can be larger depending on the label distribution of the given training data set. Furthermore, the two-qubit measurement scheme used in HTC is reduced to single-qubit measurement.

As elucidated above, the number of training data vectors in two classes may be different. In this case, the real or imaginary part of the quantum state is simply zero for the missing data. The difference in the number of training data vectors in two classes can later be compensated by controlling the weights between the real and imaginary parts of the state overlap by finding $\phi$ that satisfies
$$\frac{\sin(\phi)}{\cos(\phi)} = \frac{M_{-}}{M_{+}},$$
where $M_{\pm}$ is the number of training data points in $\pm 1$ classes.


Table~\ref{tab:accuracy} presents the classification accuracy of proof-of-concept experiments. We divided Iris and Wine datasets into training and test sets with a ratio of 80/20 and two classes and used the second and third features of the Iris dataset and the two principal components of the Wine dataset. We performed a search among 40 combinations of four training patterns (two from each class) and selected the combination with the best training accuracy. We used these four selected training points to evaluate the test accuracy. The use of two features and four training points allowed us to perform experiments in a small-scale quantum device. The accuracies and standard deviations in Table~\ref{tab:accuracy} are obtained from 30 repetitions of the random division into training and test sets. The experimental results are from \texttt{ibmq\_manila}, a quantum computer with five superconducting qubits available on the IBM quantum cloud service.

\begin{table}[h]
\begin{tabular}{l|l|l}
Dataset 				& accuracy (sim) 		& accuracy (exp) 	        \\ \hline \hline
Iris class 1{\&}2 		& 0.998 (0.009) 		& 0.952 (0.064)	        \\
Iris class 1{\&}3 		& 0.998 (0.009) 		& 0.965 (0.550) 	        \\
Iris class 2{\&}3 		& 0.922 (0.059) 		& 0.853 (0.092) 	        \\ 
Wine class 1{\&}2       & 0.930 (0.046)       & 0.894 (0.074)\\
Wine class 1{\&}3       & 0.882 (0.072)       & 0.829 (0.101)\\
Wine class 2{\&}3       & 0.694 (0.068)       & 0.657 (0.081)\\ \hline
\end{tabular}
\caption{Classification accuracy for the compact classifier on Iris dataset for 30 random separations obtained by simulation (sim) and experiment (exp). We used \texttt{ibmq\_manila}, a quantum computer made available by IBM via cloud access, to obtain the experimental (exp) results. The standard deviation appears between parenthesis.\label{tab:accuracy}}
\end{table}

\subsection{Smallest quantum binary classifier}

Let us denote $|\Psi(\mathbf{x}_j,\tilde{\mathbf{x}})\rangle = (\ket{0}\ket{\mathbf{x}_j}_c+e^{-i\phi}\ket{1}\ket{\tilde{\mathbf{x}}})/\sqrt{2}$.
By allowing classical sampling from an ensemble $\lbrace a_j , |\Psi(\mathbf{x}_j,\tilde{\mathbf{x}})\rangle\rbrace$, where $a_j$ is the probability to choose $j$th state, we can have the mixed state
\begin{equation}
    \sum_{j=0}^{\frac{M}{2}-1}a_j \ketbra{\Psi(\mathbf{x}_j,\tilde{\mathbf{x}})}{\Psi(\mathbf{x}_j,\tilde{\mathbf{x}})}.
    \label{eq:minimal}
\end{equation}
It is easy to verify that the expectation measurement of $\sigma_x$ operator (equivalent to application of a Hadamard gate followed by $\sigma_z$ measurement) yields the same outcome as shown in Eq.~(\ref{eq:expval_chtc}). In this approach, the index register is unnecessary, and hence we further reduce the number of qubits by $\lceil\log_2(M)\rceil$ and the number of gates by $O(poly(M))$. This the smallest kernel-based binary classifier; one only requires $\lceil \log_2(N)\rceil$ qubits for encoding the data set, and a qubit for measurement.

\subsection{Connection to quantum feature mapping}
Under certain restrictions, the compact encoding scheme can be applied to the quantum feature mapping framework introduced in Ref.~\cite{Havlicek2019} to store two training data in one quantum register. In principle, this can be done with the state preparation
\begin{equation}
\label{eq:featuremap}
   \ket{\Phi(\mathbf{x}_k^{\pm})}_c=\frac{ \ket{\Phi(\mathbf{x}_k^{+})}+i\ket{\Phi(\mathbf{x}_k^{-})}}{\sqrt{2}},
\end{equation}
with a feature map $U_{\Phi}(\mathbf{x})\ket{0}^{\otimes L}=\ket{\Phi(\mathbf{x})}$ that maps $N$-dimensional data to $L=O(N)$ qubits. Of course, the above state must satisfy $\langle\Phi(\mathbf{x}_k^{+})|\Phi(\mathbf{x}_k^{-})\rangle=0$. The feature map also needs to satisfy $ \braket{\Phi(\tilde{\mathbf{x}})}{\Phi(\mathbf{x}_k^{\pm})}\in\mathbb{R}$ for all $k$. One way to prepare the above state is to apply unitary transformation
$$ V = \frac{U_{\Phi}(\mathbf{x}_k^{+})+i U_{\Phi}(\mathbf{x}_k^{-})}{\sqrt{2}}$$ to $|0\rangle ^{\otimes L}$. Since $V$ is unitary, it must satisfy $U_{\Phi}(\mathbf{x}_k^{+})U_{\Phi}(\mathbf{x}_k^{-})^{\dagger}-U_{\Phi}(\mathbf{x}_k^{-})U_{\Phi}(\mathbf{x}_k^{+})^{\dagger}=0$. Finding an appropriate feature map that satisfies the above conditions while retaining the quantum advantage is a difficult task and we leave it as an interesting open problem.

Given that the state in Eq.~(\ref{eq:featuremap}) can be prepared, one can follow the strategy from the previous section. More explicitly, classical sampling from a set
$$
\left\lbrace \frac{\ket{0}\ket{\Phi(\mathbf{x}_j^{\pm})}_c  + e^{-i\phi}\ket{1}\ket{\Phi(\tilde{\mathbf{x}})}}{\sqrt{2}}\right\rbrace
$$
with probability $a_j$ followed by the expectation measurement of $\sigma_x$ operator yields
$$
\frac{1}{2}\sum_{j=0}^{\frac{M}{2}-1}a_j \left( \cos(\phi) \braket{\Phi(\tilde{\mathbf{x}})}{\Phi(\mathbf{x}_j^+)}-\sin(\phi) \braket{\Phi(\tilde{\mathbf{x}})}{\Phi(\mathbf{x}_j^-)}\right),
$$
which is reduced to the classifier of Eq.~(\ref{eq:chtc_final}) when $\phi=\pi/4$.

\section{Entanglement analysis}
\label{sec:resource}

\begin{figure*}[t]
    \centering
    \includegraphics[width=0.9\textwidth]{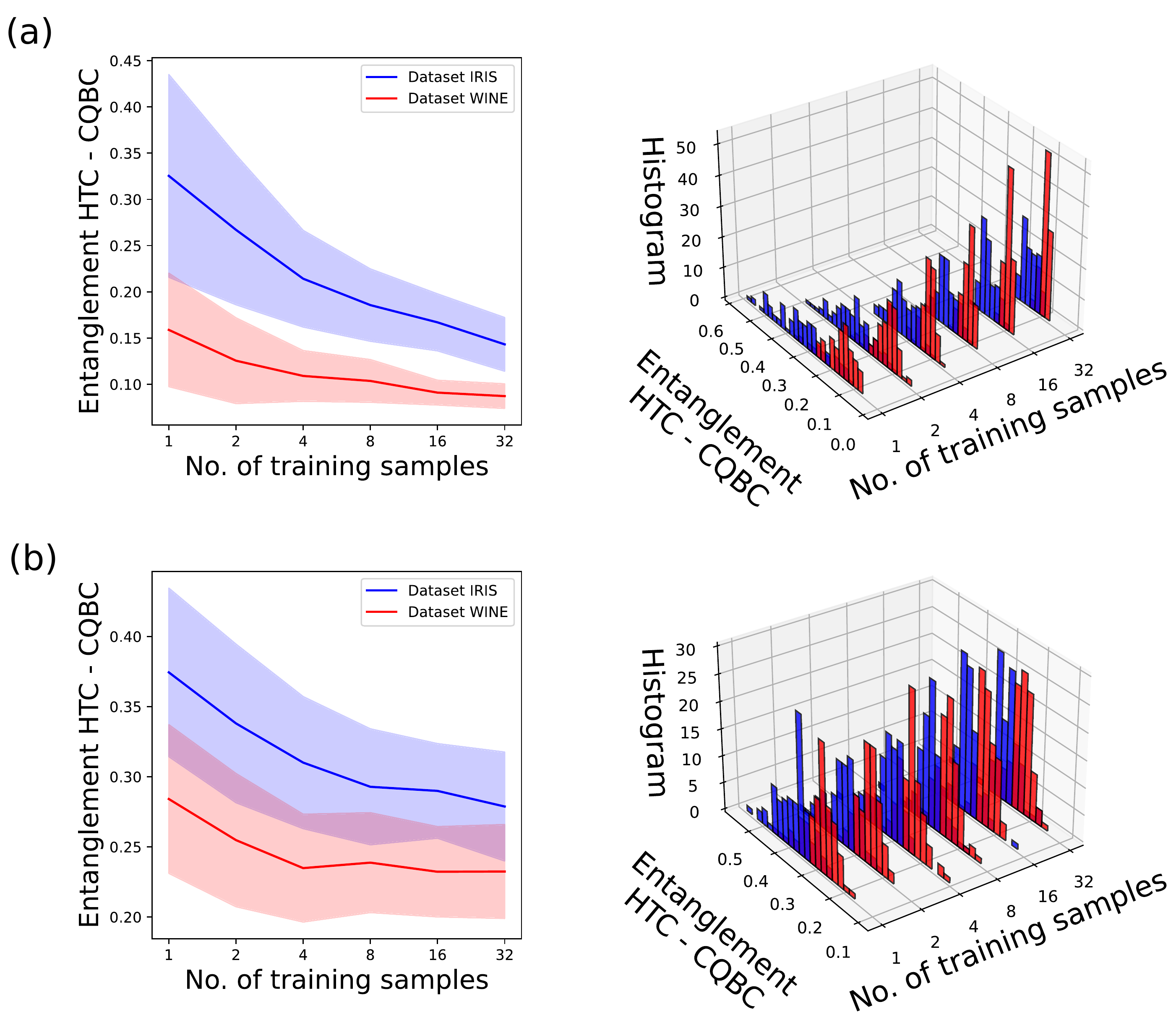}
    \caption{This figure shows the final state entanglement difference $\Delta$ of Eq.~(\ref{eq:delta_entanglement}) between the HTC and the proposed compact classifier for the (a) Meyer-Wallach and (b) the geometric measure of entanglement. It shows that for two example data sets the difference is always positive, meaning that the HTC's final state always has higher entanglement. The feature dimension is 4 for the Iris data and 13 for the Wine data.}
    \label{fig:delta_plot_entanglement}
\end{figure*}

Understanding the fundamental source of the quantum
advantage is of critical importance for establishing the
ground for further developments of new ideas. The quantum resource we considered in this Section is the entanglement of the system. Besides the fundamental perspective, entanglement is also deeply connected to the quantum circuit complexity. More specifically, it has been reported that lower amount of entanglement required in a quantum algorithm implies reduction in the number of entangling gates during state preparation~\cite{MORA2006,PhysRevLett.95.200503,2110.13454,2111.03132}.

The measure of entanglement we use in this work is the Meyer-Wallach~\cite{10.1063/1.1497700} measure, which calculates a value linearly related to the mean single-qubit purity of the state~\cite{quant-ph/0305094},
\begin{equation}
    Q^{(m)}(\ket{\psi}) = 2 \left(1 - \frac{1}{n} \sum_{k=1}^n \text{Tr}[\rho_k^2] \right)
\end{equation}
where $\rho_k$ is the single-qubit density matrix obtained by partitioning $|\psi\rangle$ into one qubit and $n-1$ qubits. The Meyer-Wallach measure is also occupied in quantifying the entangling capability of a variational quantum circuit due to its scalability and ease of computation~\cite{Sim_expressibility}. Additionally, we also apply the geometric measure of entanglement~\cite{10.1111/j.1749-6632.1995.tb39008.x,10.1103/physreva.68.042307,10.1088/0305-4470/34/35/305} with
\begin{equation}
    Q^{(g)}(\ket{\psi}) = \min_{\ket{\phi}} \| \ket{\psi} -\ket{\phi} \|
\end{equation}
where the minimization is over all states $\ket{\phi}$ that are product states, i.e., $\ket{\phi} = \otimes_{l=1}^n \ket{\phi^l}$ with each $\ket{\phi^l}$ being a local state. This can be efficiently approximated by Tucker decomposition~\cite{10.1007/s11128-017-1633-8}.

To compare the Meyer-Wallach and geometric entanglement of the final states of the CHC and HTC, we use the Iris and Wine data sets~\cite{Dua:2019}. We are interested in the value
\begin{equation}
    \label{eq:delta_entanglement}
    \Delta^{(e)} = Q^{(e)}_\text{HTC}(\ket{\psi_f}) - Q^{(e)}_\text{CHC}(\ket{\psi_f}), \quad e = m, g.
\end{equation}
The final state of each classifier is created and the measure for entanglement is calculated. The data sets used were primarily chosen for reproducibility and the different feature sizes: Iris has four features ($N=4$) and Wine has thirteen ($N=13$), for which we pad with zeros excess features due to the qubit nature. Encoding the final state, we use a number of samples of each class $M=2^m$ (for $m=0, 1, 2, 3, 4, 5$) from a test vs. train split of 2/3 train and 1/3 test chosen at random. As both data sets are not binary, we limit data to the first two classes, after all, we are not interested in the actual classification, but in the entanglement structure. The numerical survey shows that the both Meyer-Wallach {color{red}and geometrical entanglement} in the CHC is always lower than that of the HTC for all sample size $2^{m+1}$ and each data set. The results are shown in Fig.~\ref{fig:delta_plot_entanglement} and Table~\ref{tab:delta_entanglement}. In particular, the minimum $\Delta$ of each set shows that the value is always positive, hence the CHC is more resource saving in terms of entanglement. Note that there are several, non-equivalent, measures of entanglement, and a complete survey over all possible measures is beyond the scope of this work. Nevertheless, the trend we observed from evaluating two entanglement monotones advocates the compactness of the CHC.

The analysis with Iris and Wine data sets asserts that the classifier presented in this work is compact in the sense that it requires less entanglement for binary classification. The observed reduced entanglement has useful consequences. As mentioned in the beginning of this section, the lower entanglement allows for the reduction in the number of entangling gates by exploiting the consequential low Schmidt-rank of bi-partitions~\cite{2110.13454,2111.03132}. 
Furthermore, if entanglement is lower, then an approximation with even less entangling operations can be found. We suspect that this can be a useful trait for machine learning protocols, because approximation errors in the state preparation will likely only affect near-decision boundary classifications. Even though the methods described in Refs.~\cite{2110.13454,2111.03132} are computationally expensive subroutines like singular value decomposition, the reduction in the state preparation complexity can be beneficial in the NISQ era, if the classification error due to the approximation is less than that due to the hardware imperfections and decoherence.

\begin{table*}[]
    \centering
    \begin{tabular}{lrrrrrrrrrrrr}
    \toprule
    {} &  (Iris, 1) &  (Iris, 2) &  (Iris, 4) &  (Iris, 8) &  (Iris, 16) &  (Iris, 32) &  (Wine, 1) &  (Wine, 2) &  (Wine, 4) &  (Wine, 8) &  (Wine, 16) &  (Wine, 32) \\
    \midrule
    mean  &      0.251 &      0.230 &      0.186 &      0.171 &       0.157 &       0.144 &      0.064 &      0.063 &      0.062 &      0.062 &       0.059 &       0.057 \\
    std   &      0.083 &      0.061 &      0.037 &      0.026 &       0.017 &       0.016 &      0.028 &      0.020 &      0.016 &      0.011 &       0.010 &       0.008 \\
    min   &      0.072 &      0.105 &      0.111 &      0.110 &       0.114 &       0.105 &      0.020 &      0.031 &      0.038 &      0.040 &       0.045 &       0.044 \\
    25\%   &      0.190 &      0.188 &      0.159 &      0.151 &       0.146 &       0.132 &      0.043 &      0.048 &      0.050 &      0.053 &       0.052 &       0.051 \\
    50\%   &      0.256 &      0.230 &      0.180 &      0.172 &       0.160 &       0.147 &      0.060 &      0.060 &      0.060 &      0.060 &       0.057 &       0.056 \\
    75\%   &      0.312 &      0.278 &      0.211 &      0.185 &       0.167 &       0.157 &      0.078 &      0.072 &      0.071 &      0.070 &       0.065 &       0.063 \\
    max   &      0.472 &      0.420 &      0.269 &      0.237 &       0.191 &       0.168 &      0.132 &      0.129 &      0.125 &      0.088 &       0.089 &       0.074 \\
    \bottomrule
    \end{tabular}
    \caption{Data of the difference between the entanglement (Meyer-Wallach) of the HTC and the compact classifier proposed here from a statistical description of the distribution, i.e. $\Delta^{(m)} = Q^{(m)}_\text{HTC}(\ket{\psi_f}) - Q^{(m)}_\text{CQBC}(\ket{\psi_f})$. Each value is for a data type and a number of training data that are indicated in the first row as (data type, number of data). It can be seen that the minimum is always greater than zero. The Iris data set has 4 features, while the Wine data has 13 features. The table exhibits a trend that the entanglement difference decreases as the number of data register qubit increases. The table shows the statistical description using the mean, standard deviation, the minimum and maximum and finally the 25, 50, 75 percentiles.}
    \label{tab:delta_entanglement}
\end{table*}

\section{Conclusion}
\label{sec:conclusion}
This work proposes a compact quantum binary classifier whose quantum circuit size is smaller than that of the Hadamard-test classifier, which was previously considered to be the simplest kernel-based quantum classifier. Thus, our method is placed above existing kernel-based quantum classifiers as the potential application of NISQ computing. The compact quantum classifier is enabled by compact amplitude encoding we introduced in this work. This technique encodes one training data point from each class as the real and the imaginary part of the probability amplitude of a computational basis state. Since the label information of training data is implicitly encoded, there is no need  to use a separate quantum register to explicitly encode the label information as required in previous methods. The removal of the label register naturally further reduces the two-qubit measurement scheme required in previous methods  to the single-qubit measurement. Furthermore, the ease with which unbalanced data can be encoded and the applicability of feature maps is highlighted as good traits for applications. Using binary classification tasks with Iris and Wine data sets, we show that the quantum classifier proposed in this work is compact also in the sense that it requires less entanglement than the Hadamard-test classifier. 

State preparation is responsible for the main cost of CHC, HTC, and many machine learning algorithms. With a reduction in entanglement, one can attempt to create state preparation algorithms that are more efficient than the available alternatives. Reducing the circuit depth of CHC based on this intuition and the number of repetitions to generate a quantum advantage over classical algorithms for big data classification remains an important future work. Another interesting future research direction is to examine the possibility to create compact versions of other machine learning data initialization methods. For instance, some kernel methods as qubit encoding requires one qubit for each data feature. One could investigate how to reduce the required number of qubits without a reduction in the accuracy of a classifier. In addition, understanding the reason behind the reduced amount of entanglement created in the CHC algorithm compared to HTC remains an interesting open question. Answering this question could also help in designing a compact version of existing quantum machine learning algorithms. This could also lead to the discovery of quantum-inspired machine learning algorithms. Furthermore, analytical analysis of the entanglement difference ($\Delta$) for the given size of the samples ($M$) and the features ($N$) remains to be done. Another interesting question is the condition on the unitary when using quantum feature maps to encode data, both in terms of expressibility and complexity. Since the HTC and CHC construct a binary classifier based on the quantum interference effect, one can speculate that quantum coherence plays a critical role. Quantum coherence can be regarded as a resource, and it has been rigorously studied within the framework of resource theory in Refs.~\cite{PhysRevLett.113.140401,PhysRevLett.116.120404,RevModPhys.89.041003}. Connection between quantum coherence and non-classical correlation has also been studied in Refs.~\cite{PhysRevLett.116.160407,PhysRevA.97.032327}. Performing quantum resource theoretic analysis on the HTC and CHC is also remains as a potential future work.

\section*{Data availability}
The source code and data used in this study are available from the corresponding author upon reasonable request.

\section*{Acknowledgment}
This research is supported by the National Research Foundation of Korea (NRF-2019R1I1A1A01050161, NRF-2021M3H3A1038085, and NRF-2022M3E4A1074591), and the South African Research Chair Initiative of the Department of Science and Technology and the National Research Foundation.


\begin{thebibliography}{10}

\bibitem{Preskill2018quantumcomputing}
John Preskill.
\newblock Quantum {C}omputing in the {NISQ} era and beyond.
\newblock {\em {Quantum}}, 2:79, August 2018.

\bibitem{doi:10.1098/rsta.2011.0352}
Ben Criger, Gina Passante, Daniel Park, and Raymond Laflamme.
\newblock Recent advances in nuclear magnetic resonance quantum information
  processing.
\newblock {\em Philosophical Transactions of the Royal Society A: Mathematical,
  Physical and Engineering Sciences}, 370(1976):4620--4635, 2012.

\bibitem{doi:https://doi.org/10.1002/9781118742631.ch08}
Ben Criger, Daniel Park, and Jonathan Baugh.
\newblock {\em Few-Qubit Magnetic Resonance Quantum Information Processors:
  Simulating Chemistry and Physics}, pages 193--228.
\newblock John Wiley \& Sons, Ltd, 2014.

\bibitem{Park_Kyungdeock2015}
{Park, Kyungdeock}.
\newblock {\em Coherent control of nuclear and electron spins for quantum
  information processing}.
\newblock PhD thesis, 2015.

\bibitem{Google_QS}
Frank Arute, , et~al.
\newblock Quantum supremacy using a programmable superconducting processor.
\newblock {\em Nature}, 574(7779):505--510, 2019.

\bibitem{ion_trap_BM}
K.~Wright et~al.
\newblock Benchmarking an 11-qubit quantum computer.
\newblock {\em Nature Communications}, 10(1):5464, 2019.

\bibitem{HF_google}
Frank Arute, , et~al.
\newblock Hartree-fock on a superconducting qubit quantum computer.
\newblock {\em Science}, 369(6507):1084--1089, 2020.

\bibitem{Zhong1460}
Han-Sen Zhong et~al.
\newblock Quantum computational advantage using photons.
\newblock {\em Science}, 370(6523):1460--1463, 2020.

\bibitem{harrigan_quantum_2021}
Matthew~P. Harrigan et~al.
\newblock Quantum approximate optimization of non-planar graph problems on a
  planar superconducting processor.
\newblock {\em Nature Physics}, February 2021.

\bibitem{wittek}
Peter Wittek.
\newblock {\em Quantum Machine Learning: What Quantum Computing Means to Data
  Mining}.
\newblock Academic Press, Boston, 2014.

\bibitem{QML-Biamonte}
Jacob Biamonte, Peter Wittek, Nicola Pancotti, Patrick Rebentrost, Nathan
  Wiebe, and Seth Lloyd.
\newblock Quantum machine learning.
\newblock {\em Nature}, 549:195--202, 2017.

\bibitem{SupervisedQML}
Maria Schuld and Francesco Petruccione.
\newblock {\em Supervised Learning with Quantum Computers}.
\newblock Quantum Science and Technology. Springer International Publishing,
  2018.

\bibitem{Dunjko_2018}
Vedran Dunjko and Hans~J Briegel.
\newblock Machine learning {\&} artificial intelligence in the quantum domain:
  a review of recent progress.
\newblock {\em Reports on Progress in Physics}, 81(7):074001, jun 2018.

\bibitem{QML_PRSA}
Carlo Ciliberto, Mark Herbster, Alessandro~Davide Ialongo, Massimiliano Pontil,
  Andrea Rocchetto, Simone Severini, and Leonard Wossnig.
\newblock Quantum machine learning: a classical perspective.
\newblock {\em Proceedings of the Royal Society A: Mathematical, Physical and
  Engineering Sciences}, 474(2209):20170551, 2018.

\bibitem{schuld2021quantum}
Maria Schuld.
\newblock Supervised quantum machine learning models are kernel methods.
\newblock {\em arXiv preprint arXiv:2101.11020}, 2021.

\bibitem{PhysRevLett.100.160501}
Vittorio Giovannetti, Seth Lloyd, and Lorenzo Maccone.
\newblock Quantum random access memory.
\newblock {\em Phys. Rev. Lett.}, 100:160501, Apr 2008.

\bibitem{ffqram}
Daniel~K. Park, Francesco Petruccione, and June-Koo~Kevin Rhee.
\newblock Circuit-based quantum random access memory for classical data.
\newblock {\em Scientific Reports}, 9(1):3949, 2019.

\bibitem{9259210}
Tiago M.~L. de~Veras, Ismael C.~S. de~Araujo, Daniel~K. Park, and Adenilton~J.
  da~Silva.
\newblock Circuit-based quantum random access memory for classical data with
  continuous amplitudes.
\newblock {\em IEEE Transactions on Computers}, 70(12):2125--2135, 2021.

\bibitem{Nielsen:2011:QCQ:1972505}
Michael~A. Nielsen and Isaac~L. Chuang.
\newblock {\em Quantum Computation and Quantum Information: 10th Anniversary
  Edition}.
\newblock Cambridge University Press, New York, NY, USA, 10th edition, 2011.

\bibitem{PhysRevLett.103.150502_HHL_qBLAS}
Aram~W. Harrow, Avinatan Hassidim, and Seth Lloyd.
\newblock Quantum algorithm for linear systems of equations.
\newblock {\em Phys. Rev. Lett.}, 103:150502, Oct 2009.

\bibitem{PhysRevLett.113.130503_qSQVM}
Patrick Rebentrost, Masoud Mohseni, and Seth Lloyd.
\newblock Quantum support vector machine for big data classification.
\newblock {\em Phys. Rev. Lett.}, 113:130503, Sep 2014.

\bibitem{qPCA}
Seth Lloyd, Masoud Mohseni, and Patrick Rebentrost.
\newblock Quantum principal component analysis.
\newblock {\em Nature Physics}, 10(9):631--633, 2014.

\bibitem{Havlicek2019}
Vojtech Havl{\'i}cek, Antonio~D. C{\'o}rcoles, Kristan Temme, Aram~W. Harrow,
  Abhinav Kandala, Jerry~M. Chow, and Jay~M. Gambetta.
\newblock Supervised learning with quantum-enhanced feature spaces.
\newblock {\em Nature}, 567(7747):209--212, 2019.

\bibitem{PhysRevLett.122.040504}
Maria Schuld and Nathan Killoran.
\newblock Quantum machine learning in feature hilbert spaces.
\newblock {\em Phys. Rev. Lett.}, 122:040504, Feb 2019.

\bibitem{htc}
M.~Schuld, M.~Fingerhuth, and F.~Petruccione.
\newblock Implementing a distance-based classifier with a quantum interference
  circuit.
\newblock {\em EPL (Europhysics Letters)}, 119(6):60002, 2017.

\bibitem{blank_quantum_2020}
Carsten Blank, Daniel~K. Park, June-Koo~Kevin Rhee, and Francesco Petruccione.
\newblock Quantum classifier with tailored quantum kernel.
\newblock {\em npj Quantum Information}, 6(1):41, May 2020.

\bibitem{PARK2020126422}
Daniel~K. Park, Carsten Blank, and Francesco Petruccione.
\newblock The theory of the quantum kernel-based binary classifier.
\newblock {\em Physics Letters A}, 384(21):126422, 2020.

\bibitem{MORA2006}
Caterina~E. Mora and Hans~J. Briegel.
\newblock Algorithmic complexity of quantum states.
\newblock {\em International Journal of Quantum Information}, 04:715--737,
  2006.

\bibitem{PhysRevLett.95.200503}
Caterina~E. Mora and Hans~J. Briegel.
\newblock Algorithmic complexity and entanglement of quantum states.
\newblock {\em Phys. Rev. Lett.}, 95:200503, 2005.

\bibitem{2110.13454}
Prithvi Gundlapalli and Junyi Lee.
\newblock Deterministic, scalable, and entanglement efficient initialization of
  arbitrary quantum states.
\newblock {\em arXiv preprint arXiv:2110.13454}, 2021.

\bibitem{2111.03132}
Israel~F Araujo, Carsten Blank, and Adenilton~J da~Silva.
\newblock Entanglement as a complexity measure for quantum state preparation.
\newblock {\em arXiv preprint arXiv:2111.03132}, 2021.

\bibitem{giuntini2021quantum}
Roberto Giuntini, Hector Freytes, Daniel~K Park, Carsten Blank, Federico Holik,
  Keng~Loon Chow, and Giuseppe Sergioli.
\newblock Quantum state discrimination for supervised classification.
\newblock {\em arXiv preprint arXiv:2104.00971}, 2021.

\bibitem{qcnn}
Tak Hur, Leeseok Kim, and Daniel~K. Park.
\newblock Quantum convolutional neural network for classical data
  classification.
\newblock {\em Quantum Machine Intelligence}, 4(1):3, 2022.

\bibitem{park2021robust}
Daniel~K. Park, Carsten Blank, and Francesco Petruccione.
\newblock Robust quantum classifier with minimal overhead.
\newblock In {\em 2021 International Joint Conference on Neural Networks
  (IJCNN)}, pages 1--7, 2021.

\bibitem{Mottonen:2005:TQS:2011670.2011675}
Mikko M\"{o}tt\"{o}nen, Juha~J. Vartiainen, Ville Bergholm, and Martti~M.
  Salomaa.
\newblock Transformation of quantum states using uniformly controlled
  rotations.
\newblock {\em Quantum Info. Comput.}, 5(6):467--473, September 2005.

\bibitem{bergholm2005quantum}
Ville Bergholm, Juha~J Vartiainen, Mikko M{\"o}tt{\"o}nen, and Martti~M
  Salomaa.
\newblock Quantum circuits with uniformly controlled one-qubit gates.
\newblock {\em Physical Review A}, 71(5):052330, 2005.

\bibitem{PhysRevA.52.3457}
Adriano Barenco, Charles~H. Bennett, Richard Cleve, David~P. DiVincenzo, Norman
  Margolus, Peter Shor, Tycho Sleator, John~A. Smolin, and Harald Weinfurter.
\newblock Elementary gates for quantum computation.
\newblock {\em Phys. Rev. A}, 52:3457--3467, Nov 1995.

\bibitem{10.1063/1.1497700}
David~A. Meyer and Nolan~R. Wallach.
\newblock {Global entanglement in multiparticle systems}.
\newblock {\em Journal of Mathematical Physics}, 43(9):4273--4278, 2002.

\bibitem{quant-ph/0305094}
Gavin~K Brennen.
\newblock An observable measure of entanglement for pure states of multi-qubit
  systems.
\newblock {\em arXiv preprint quant-ph/0305094}, 2003.

\bibitem{Sim_expressibility}
Sukin Sim, Peter~D. Johnson, and Alán Aspuru-Guzik.
\newblock Expressibility and entangling capability of parameterized quantum
  circuits for hybrid quantum-classical algorithms.
\newblock {\em Advanced Quantum Technologies}, 2(12):1900070, 2019.

\bibitem{10.1111/j.1749-6632.1995.tb39008.x}
ABNER SHIMONY.
\newblock {Degree of Entanglementa}.
\newblock {\em Annals of the New York Academy of Sciences}, 755(1):675--679,
  1995.

\bibitem{10.1103/physreva.68.042307}
Tzu-Chieh Wei and Paul~M. Goldbart.
\newblock {Geometric measure of entanglement and applications to bipartite and
  multipartite quantum states}.
\newblock {\em Physical Review A}, 68(4):042307, 2003.

\bibitem{10.1088/0305-4470/34/35/305}
H~Barnum and N~Linden.
\newblock {Monotones and invariants for multi-particle quantum states}.
\newblock {\em Journal of Physics A: Mathematical and General}, 34(35):6787,
  2001.

\bibitem{10.1007/s11128-017-1633-8}
Peiyuan Teng.
\newblock {Accurate calculation of the geometric measure of entanglement for
  multipartite quantum states}.
\newblock {\em Quantum Information Processing}, 16(7):181, 2017.

\bibitem{Dua:2019}
Dheeru Dua and Casey Graff.
\newblock {UCI} machine learning repository, 2017.

\bibitem{PhysRevLett.113.140401}
T.~Baumgratz, M.~Cramer, and M.~B. Plenio.
\newblock Quantifying coherence.
\newblock {\em Phys. Rev. Lett.}, 113:140401, Sep 2014.

\bibitem{PhysRevLett.116.120404}
Andreas Winter and Dong Yang.
\newblock Operational resource theory of coherence.
\newblock {\em Phys. Rev. Lett.}, 116:120404, Mar 2016.

\bibitem{RevModPhys.89.041003}
Alexander Streltsov, Gerardo Adesso, and Martin~B. Plenio.
\newblock Colloquium: Quantum coherence as a resource.
\newblock {\em Rev. Mod. Phys.}, 89:041003, Oct 2017.

\bibitem{PhysRevLett.116.160407}
Jiajun Ma, Benjamin Yadin, Davide Girolami, Vlatko Vedral, and Mile Gu.
\newblock Converting coherence to quantum correlations.
\newblock {\em Phys. Rev. Lett.}, 116:160407, Apr 2016.

\bibitem{PhysRevA.97.032327}
Daniel~K. Park, June-Koo~K. Rhee, and Soonchil Lee.
\newblock Noise-tolerant parity learning with one quantum bit.
\newblock {\em Phys. Rev. A}, 97:032327, Mar 2018.

\end{thebibliography}

\end{document}